\documentclass[12pt]{article}
\usepackage[latin1]{inputenc}
\usepackage{pstricks}
\usepackage{amsmath}
\usepackage{float}
\usepackage{color}
\input epsf
\usepackage{amssymb}
\usepackage[dvips]{graphicx,psfrag}
\newcommand{\be}{\begin{equation}}
\newcommand{\ee}{\end{equation}}
\newcommand{\bea}{\begin{eqnarray}}
\newcommand{\eea}{\end{eqnarray}}
\newcommand{\lb}{\label}
\begin{document}
\begin{titlepage}
\title{The growth of matter perturbations in f(R) models}
\author{R. Gannouji\thanks{email:gannouji@lpta.univ-montp2.fr}~,
B. Moraes\thanks{email:moraes@lpta.univ-montp2.fr} and  
D. Polarski\thanks{email:polarski@lpta.univ-montp2.fr}\\
\hfill\\
Lab. de Physique Th\'eorique et Astroparticules, CNRS\\ 
Universit\'e Montpellier II, France}
\pagestyle{plain}
\date{\today}

\maketitle

\begin{abstract}
We consider the linear growth of matter perturbations on low redshifts 
in some $f(R)$ dark energy (DE) models. We discuss the definition of dark 
energy (DE) in these models and show the differences with scalar-tensor DE 
models. For the $f(R)$ model recently proposed by Starobinsky we show that 
the growth parameter $\gamma_0\equiv \gamma(z=0)$ takes the value 
$\gamma_0\simeq 0.4$ for $\Omega_{m,0}=0.32$ and $\gamma_0\simeq 0.43$ 
for $\Omega_{m,0}=0.23$, allowing for a clear distinction from $\Lambda$CDM. 
Though a scale-dependence appears in the growth of perturbations on higher 
redshifts, we find no dispersion for $\gamma(z)$ on low redshifts up to 
$z\sim 0.3$, $\gamma(z)$ is also quasi-linear in this interval. At redshift 
$z=0.5$, the dispersion is still small with  $\Delta \gamma\simeq 0.01$.
As for some scalar-tensor models, we find here too a large value 
for $\gamma'_0\equiv \frac{d\gamma}{dz}(z=0)$, $\gamma'_0\simeq -0.25$ 
for $\Omega_{m,0}=0.32$ and $\gamma'_0\simeq -0.18$ for $\Omega_{m,0}=0.23$.  
These values are largely outside the range found for DE models in General 
Relativity (GR). This clear signature provides a powerful constraint on 
these models. 
\end{abstract}

PACS Numbers: 04.62.+v, 98.80.Cq
\end{titlepage}

\section{Introduction}
The present stage of accelerated expansion of the universe  \cite{P97} is a major challenge 
for cosmology. There are basically two main roads one can take: either to introduce a new 
(approximately) smooth component or to modify the laws of gravity on cosmic scales. 
In the first class of models, the new component dubbed dark energy (DE) must have a 
sufficiently negative pressure in order to induce a late-time accelerated expansion. 
One typically adds a new isotropic perfect fluid with negative pressure able to 
accelerate the expansion. In the second class of models one is trying to explain the 
accelerated expansion by modifying gravity, after all the universe expansion is a 
large-scale gravitational effect. It is far from clear at the present time which class 
of DE models will finally prevail and one must study carefully all possibilities. 
While all models are called DE models \cite{SS00}, the models of the second class are 
called more specifically modified gravity DE models.

It is clear that in both classes we have many models able to reproduce a late-time 
accelerated expansion in agreement with the present data probing the background 
expansion like luminosity distances from SNIa. Probably this will remain the case even 
with more accurate data. However the evolution of matter perturbations will be 
affected differently depending on the class of models we are dealing with. This can 
be used for a consistency check in order to find out whether a DE model is inside 
General Relativity (GR) or not \cite{S98}. Therefore the evolution of matter 
perturbations seems to be a powerful tool to discriminate between models inside and 
outside GR \cite{B06}. In particular, the growth of matter perturbations on small 
redshifts seems a promising tool and it has been the subject of intensive 
investigations recently \cite{HL06}. 

A particular class of modified gravity DE models are $f(R)$ models where $R$ is 
replaced by some function $f(R)$ in the gravitational Lagrangian density \cite{Cap02} 
(for a recent review, see \cite{SF08}). 
The idea of producing an accelerated stage in such models was first successfully 
implemented in Starobinsky's inflationary model \cite{S80}. In some sense it was 
rather natural to try explaining the late-time accelerated expansion by the same 
kind of mechanism, a modification of gravity but now at much lower energies. Some 
problems arising in these $f(R)$ DE models were quickly pointed out, related to 
solar-system constraints \cite{Chi03} and instabilities \cite{Dol03}.  
It was then found that a very serious problem arises where it was not expected:  
many of these models are unable to produce the late-time acceleration together with 
a viable cosmic expansion history \cite{APT07}. It was shown that many popular $f(R)$ 
models, like those containing a $R^{-1}$ term \cite{CDTT04}, are unable to account for 
a viable expansion history because of the absence of a standard matter-dominated stage 
$a\propto t^{\frac{2}{3}}$ which is replaced instead by the behaviour 
$a\propto t^{\frac{1}{2}}$ \cite{AGPT07}. 

However some interesting cosmological models remain viable like the model recently 
suggested by Starobinsky \cite{S07}, (for another interesting viable model see also 
\cite{HS07}). 
It is interesting in the first place because of its 
non-trivial ability to allow for a standard matter-dominated stage. This goes together 
with the appearance of large oscillations in the past and the overproduction 
of massive scalar particles in the very early universe which, as noted already in 
\cite{S07}, can be a serious problem of all viable $f(R)$ DE models. Another interesting 
property is the scale dependence of the growth of matter perturbations deep inside the 
Hubble radius \cite{T07},\cite{B07}. This scale dependence was actually used in \cite{S07} 
in order to constrain one of the parameters of the model. Of course this model also 
satisfies the local gravity constraints for certain parameter values. In the end a window 
in parameter space remains where the model is in principle physically acceptable. 
    
In this letter we want to assess the viability of this model with respect to the growth 
of matter perturbations (see e.g. \cite{DJJNST08} for other possible constraints and 
approaches). Some 
viable $f(R)$ DE models have their free parameters constrained in such a way that they 
cannot be distinguished observationally from $\Lambda$CDM. 
The situation is different in this case. Our results show that the growth of matter 
perturbations on low redshifts $z\lesssim 1$ provides a discriminating signature 
of these models, able to clearly differentiate it from $\Lambda$CDM and also from any 
other DE model inside GR. These results confirm that the growth of matter perturbations 
can be used efficiently to track the nature of DE models.  
%%%%%%%%%%%%%%%%%%%%%%%%%%%%%%%%%%%%%%%%%%%%%%%%%%%%%%%%%%%%%%%%%%%%%%%%%%%%%%%%%%%%%%%%%%%
%%%%%%%%%%%%%%%%%%%%%%%%%%%%%%%%%%%%%%%%%%%%%%%%%%%%%%%%%%%%%%%%%%%%%%%%%%%%%%%%%%%%%%%%%%%
%%%%%%%%%%%%%%%%%%%%%%%%%%%%%%%%%%%%%%%%%%%%%%%%%%%%%%%%%%%%%%%%%%%%%%%%%%%%%%%%%%%%%%%%%%%

\section{$f(R)$ modified gravity cosmology}

We consider now $f(R)$ models and discuss some general properties in connection 
with their observational viability. Many $f(R)$ models have the surprising 
property that they cannot account simultaneously for a viable cosmic expansion 
history and a late-time accelerated expansion. This severely constraints viable 
$f(R)$ modified gravity DE models. We consider a universe where gravity and the 
content of the universe are described by the following action
\be
\mathcal{S}~=~\int{d^4x}\sqrt{-g}\left[\frac{1}{16\pi G_*}f(R) + \mathcal{L}_m\right]~.
\ee
For the time being, $G_*$ is a bare gravitational constant, its connection with 
observations depends on the theory under consideration. 
We concentrate on spatially flat Friedman-Lema\^{\i}tre-Robertson-Walker
(FLRW) universes with a time-dependent scale factor $a(t)$ and a
metric 
\be
ds^{2}=-dt^{2}+a^{2}(t)~d{\bf x}^{2}~.
\ee 
For this metric the Ricci scalar $R$ is given by 
\be 
R=6\left(2H^{2}+\dot{H}\right)~,\label{R}
\ee
where $H\equiv \frac{\dot{a}}{a}$ is the Hubble rate while a dot stands for
a derivative with respect to the cosmic time $t$. DE models have $\ddot{a}>0$ 
on low redshifts. The following equations are obtained 
\bea
3FH^{2} & = & 8\pi G_* ~(\rho_m +\rho_{rad})+\frac{1}{2}(FR-f)-3H\dot{F}~,\label{E1}\\
-2F\dot{H} & = & 8\pi G_* \left( \rho_m + \frac{4}{3}\rho_{rad} \right)+ 
                                      \ddot{F}-H\dot{F}~,\label{E2}
\eea
where 
\be 
F\equiv\frac{df}{dR}~.
\ee 
In standard Einstein gravity ($f=R$) one has $F=1$. The densities $\rho_{{\rm m}}$ 
and $\rho_{{\rm rad}}$ satisfy the usual conservation equations 
$\dot{\rho}_i= - 3H (1 + w_i) \rho_i$ with $w_m=0$ and $w_{rad}= \frac{1}{3}$. 

We summarize briefly the way in which the dynamics of $f(R)$ models can 
be analyzed. For a general $f(R)$ model we introduce the following 
(dimensionless) variables 
\bea 
x_{1} & = & -\frac{\dot{F}}{HF}\,,\label{x1}\\
x_{2} & = & -\frac{f}{6FH^{2}}\,,\label{x2}\\
x_{3} & = & \frac{R}{6H^{2}}=\frac{\dot{H}}{H^{2}}+2\,,\label{x3}\\
x_{4} & = & ~\frac{8\pi G_* \rho_{rad}}{3FH^2}\,.\label{x4}
\eea
From Eq.~(\ref{E1}) we have the algebraic identity 
\be
\tilde{\Omega}_m = \frac{8\pi G_*~\rho_m}{3FH^{2}}=1 - x_1 - x_2 - x_3 - x_4~.\lb{Omem}
\ee 
It is then possible to write down the following autonomous system  
\bea
\frac{dx_{1}}{dN} & = & -1-x_{3}-3x_{2}+x_{1}^{2}-x_{1}x_{3}+x_{4}~,\label{N1}\\
\frac{dx_{2}}{dN} & = & \frac{x_{1}x_{3}}{m}-x_{2}(2x_{3}-4-x_{1})~,\label{N2}\\
\frac{dx_{3}}{dN} & = & -\frac{x_{1}x_{3}}{m}-2x_{3}(x_{3}-2)~,\label{N3}\\
\frac{dx_{4}}{dN} & = & -2x_{3}x_{4}+x_{1}\, x_{4}\,,\label{N4}
\eea
where $N\equiv \ln a$. In eqs(\ref{N2},\ref{N3}), the quantity $m$ corresponds to 
\be
m \equiv \frac{d \log F}{d \log R}=\frac{R F'}{F}~,\lb{mdef}
\ee
(a prime stands for derivative with respect to $R$) hence $m$ is a priori a function 
of $R$. However it is easy to see that $R$ can in turn be expressed in function of 
the variables of our autonomous system. Indeed we have
\be
\frac{x_3}{x_2} \equiv r = -\frac{R F}{f} = -\frac{d \log f}{d \log R}~.\lb{ldef}
\ee
Inverting (\ref{ldef}) we can in principle express $R$ as a function of 
$\frac{x_3}{x_2}\equiv r$ and so $m$ becomes in turn a function of $r,~m=m(r)$ 
which closes our system. 

It was found in \cite{AGPT07} that when radiation is negligible ($x_4=0$) there 
exists a critical point $x_1=0, x_2=-\frac{1}{2}, x_3=\frac{1}{2}$ corresponding 
to an exact matter phase 
\bea 
w_{\rm eff} &=& - 1 - \frac{2\dot{H}}{3H^{2}} = \frac{1}{3}(1 - 2x_3)=0\\
\tilde{\Omega}_m &=& 1~.
\eea
This is the critical point called $P_5(m=0)$ in \cite{AGPT07}, it 
satisfies $r=-1$. 

Let us comment about the physical meaning of these conditions. 
For $r=-1$ or $x_2 + x_3 = 0$, eq.(\ref{E1}) reduces to 
\be
3FH^{2} = 8\pi G_*~\rho_m - 3H\dot{F}~.
\ee
Then we clearly have an exact matter phase ($a\sim t^{\frac{2}{3}}$) 
for $\dot{F} = 0$ or $x_1=0$. 
We note further from eqs.(\ref{N2},\ref{N3}) that $x_2+x_3$ remains 
zero for $x_1=0$.

The condition $m=0$ is more subtle. From eq.(\ref{N3}) a stable matter 
phase requires 
\be
x_1 = 3~m~.\lb{x1m}
\ee  
It is straightforward to generalize (\ref{x1m}) for arbitrary scaling 
behaviour of the background 
\be
x_1 = 3~m (1 + w_{\rm eff})~.\lb{x1mb}
\ee
Actually neither $x_1$ nor $m$ can be exactly zero during a matter 
phase (which corresponds to a critical point of the system). This 
would require $F'=0$, which reduces to General Relativity plus a 
cosmological constant. 
The system can only come in the vicinity of $P_5$ and trajectories 
with an acceptable matter phase have $x_1\approx 0$ and $m\approx 0$ with 
$x_1\approx 3~m$. In addition, from $F'>0$ both $x_1$ and $m$ should be 
positive during the matter phase.
Finally, it is easy to check that (\ref{x1m}) implies 
\be
R\propto a^{-3}~.
\ee
The situation is very different in scalar-tensor models where a standard matter 
phase is possible with $\dot{F}\approx 0$ because the energy density associated 
with the dilaton $\phi$ can mimic the behaviour of dustlike matter for a 
particular type of potential $U(\phi)$ \cite{GPRS06}. It is this property which 
is used in \cite{GP08}. 
More explicitly, for the Lagrangian density 
\begin{equation}
{\cal L} = \frac{1}{16\pi G_*} \Bigl( F(\Phi)~R - g^{\mu\nu}\partial_{\mu}\Phi\partial_{\nu}
\Phi - 2U(\Phi) \Bigr) + L_m(g_{\mu\nu})~,
\label{L}
\end{equation}
we have the Friedmann equations (in flat space)
\bea
3FH^2 &=& 8 \pi G_*~\rho_m + \frac{\dot{\Phi}^2}{2} + U - 3H {\dot F}~,\lb{E1st}\\
-2 F {\dot H} &=& 8 \pi G_*~\rho_m + \dot{\Phi}^2 + {\ddot F} - H {\dot F}~.\lb{E2st}
\eea
It is possible to have a solution with ${\dot F}=0$ and $\frac{\dot{\Phi}^2}{2} + U\propto 
a^{-3}$ by a suitable choice of the potential $U$ \cite{GPRS06}. Because of this property 
it is possible in the matter phase to have a nonvanishing DE energy density $\rho_{DE}$. 
The definition of $\rho_{DE}$ requires some care. If we want $\rho_{DE}$ to obey the usual 
energy conservation for perfect fluids, certainly a desirable property, then the choice is 
not unique. The Friedmann equations (\ref{E1},\ref{E2}) and also
(\ref{E1st},\ref{E2st}) can be recast in the form
\bea
3 A~H^2 &=& 8 \pi G_*~\left(\rho_m + \rho_{DE}(A) \right) \lb{E1A} \\
-2A~{\dot H} &=& 8 \pi G_*~\left( \rho_m + \rho_{DE}(A) + p_{DE}(A) \right) ~. \lb{E2A}
\eea
where $A$ is some arbitrary constant. We have written $\rho_{DE}(A)$ (and $p_{DE}(A)$) 
as these quantities depend on the choice of $A$. 
%%%%%%%%%%%%%%%%%%%%%
%%%%%%%%%%%%%%%%%%%%%
For example from (\ref{E1},\ref{E2}), neglecting radiation, we obtain
\bea
8 \pi G_*~\rho_{DE}(A) &=& \frac{1}{2}(FR-f)-3H\dot{F} + 3 (A - F)~H^2 \\
8 \pi G_*~\left( \rho_{DE}(A) + p_{DE}(A) \right) &=& \ddot{F}-H\dot{F} -2(A-F)~{\dot H}
\eea
%%%%%%%%%%%%%%%%%%%%%
%%%%%%%%%%%%%%%%%%%%%
Obviously, whatever choice is adopted the corresponding DE component obeys the 
usual conservation law of an isotropic perfect fluid. 
For the representation (\ref{E1A}), it is natural to introduce the cosmic relative 
densities 
\be
\Omega_i\equiv \frac{8 \pi G_* \rho_i}{3A H^2}~.\lb{OmA}
\ee
A natural choice, especially if we are willing to compare modified 
gravity theories with General Relativity (GR), is to take  
\be
\frac{G_*}{A} \simeq G_N~,
\ee
to very high accuracy, where $G_N$ is Newton's constant found in textbooks. In GR, 
it is the bare gravitational constant appearing in the action and it corresponds 
to the value of the gravitational constant obtained in a Cavendish-type experiment 
measuring the Newtonian force between two close test masses. As is well-known, this 
identification no longer holds outside GR. For modified gravity theories it is the 
quantity $G_{\rm eff}$, a priori different from $G_*$, which plays the role of 
gravitational constant and which is measured in a Cavendish-type experiment and 
clearly all acceptable models must have the same $G_{\rm eff}$ today to very high 
precision. 

For scalar-tensor models we can make the choice $A=F_0$ \cite{GPRS06} as 
$\frac{G_*}{F_0}\simeq G_{{\rm eff},0}$ to very high accuracy because 
$\omega_{BD,0}>4\times 10^4$ from solar system constraints. 
As we will see later, in our $f(R)$ model a Cavendish-type experiment corresponds to 
the limit $G_{{\rm eff}}=G_*$ (see \cite{S07} and eq. \eqref{Vr} below). It is 
therefore natural to take $A=1$ in such a model. 

As noted in \cite{S07} a consequence of the arbitrariness in the choice of $A$ is that 
it introduces an arbitrariness in the sign of $\rho_{DE}$ as well.
In a model where $F_0>F$ in the past, $\rho_{DE}$ remains positive if we take $A=F_0$. 
In our $f(R)$ models however, we have $F_0<F(z>0)$ in an expanding universe with 
$F\to 1$ at high redshifts. Hence $\rho_{DE}(A)$ defined from (\ref{E1A}) can become 
negative in the past if we take $A=F_0$. This problem is avoided if we define $\rho_{DE}$ 
from (\ref{E1A},\ref{E2A}) putting $A=1$ 
\bea
3H^2 &=& 8 \pi G_*~\left(\rho_m + \rho_{DE} \right) \lb{E1fR} \\
-2{\dot H} &=& 8 \pi G_*~\left( \rho_m + \rho_{DE} + p_{DE} \right) ~. \lb{E2fR}
\eea
The corresponding relative densities become $\Omega_i\equiv \frac{8 \pi G_* \rho_i}{3 H^2}$~
and actually $\Omega_{DE}\to 0$ in the past. 
Note that though it is always possible to write the Friedmann equations in the form 
(\ref{E1fR},\ref{E2fR}) this should not conceal the fact that we are dealing with a 
genuinely modified gravity theory obeying eqs.(\ref{E1},\ref{E2}). The difference is 
absorbed in the microscopic definitions of the DE component. This shows again the 
need to go beyond the background expansion in order to discriminate modified gravity 
DE models from GR. 
%%%%%%%%%%%%%%%%%%%%%%%%%%%%%%%%%%%%%%%%%%%%%%%%%%%%%%%%%%%%%%%%%%%%%%%%%%%%%%%%%%%%%%%%%%%
%%%%%%%%%%%%%%%%%%%%%%%%%%%%%%%%%%%%%%%%%%%%%%%%%%%%%%%%%%%%%%%%%%%%%%%%%%%%%%%%%%%%%%%%%%%
%%%%%%%%%%%%%%%%%%%%%%%%%%%%%%%%%%%%%%%%%%%%%%%%%%%%%%%%%%%%%%%%%%%%%%%%%%%%%%%%%%%%%%%%%%%
%\section{Some f(R) models}
\section{Starobinsky's model}
It was shown that a viable cosmic expansion is rather problematic in $f(R)$ models. 
%%%%%%%%%%%%%%%%%%%%%%%%%%%%%%%%%%%%%%%%%%%%%%%%%%%%%%%%%
%%%%%%%%%%%%%%%%%%%%%%%%%%%%%%%%%%%%%%%%%%%%%%%%%%%%%%%%%
As already mentioned in the Introduction problems related to instabilities \cite{Dol03} and 
solar-system constraints \cite{Chi03} were pointed out soon after the first attempts to 
produce $f(R)$ DE models. An even more unexpected problem was found later related to the background 
expansion as many $f(R)$ models, including popular ones containing an inverse power of $R$, are 
unable to reproduce a viable expansion history containing a standard matter-dominated stage 
\cite{APT07}. The viability of these models was then systematically studied in \cite{AGPT07}. 
All this shows that the construction of a viable $f(R)$ DE model is something highly non-trivial. 
%%%%%%%%%%%%%%%%%%%%%%%%%%%%%%%%%%%%%%%%%%%%%%%%%%%%%%%%%
%%%%%%%%%%%%%%%%%%%%%%%%%%%%%%%%%%%%%%%%%%%%%%%%%%%%%%%%%
Still there are some $f(R)$ DE models left that can account for a viable expansion 
history and still depart from $\Lambda$CDM. One such interesting model was suggested 
recently by Starobinsky
\be
f(R)~=~R+\lambda R_c \left(\left(1+\frac{R^2}{R_c^2}\right)^{-n}-1\right)~,
\ee
with $\lambda,~n>0$. We see that $f(R)\to R$ for $R\to 0$ in contrast to models 
containing an inverse power of $R$, in addition in flat space the correction 
term to Einstein gravity disappears. 
%%%%%%%%%%%%%%%%%%%%%%%%
%%%%%%%%%%%%%%%%%%%%%%%%
The quantity $R_c$ has the dimension of $R$ and is a free parameter 
of the model, it corresponds essentially to the present cosmic value of $R$. 
Recently problems related to the high-curvature regime of this model were pointed 
out \cite{Fro08} but it seems this could be cured by the addition of a 
term $\frac{R^2}{\overline{m}^2}$ as noted in \cite{S07}. 
However the mass $\overline{m}$ is required to be very high with $\overline{m}\gg 1$ GeV 
hence this term is negligible in the low curvature regime. 
In the present work 
we will show that it is the low curvature limit of this model, through the 
growth of matter perturbations on small redshifts, that could severely constrain 
this model. 
%%%%%%%%%%%%%%%%%%%%%%%%
%%%%%%%%%%%%%%%%%%%%%%%%
It is convenient to introduce the reduced dimensionless curvature 
$\tilde{R}\equiv \frac{R}{R_c}$.  
We have for this model
\be
r(R)=\frac{2 n  \lambda \tilde{R} \left( \tilde{R}^2 + 1 \right)^{-n-1}-1}
                {1+\lambda \tilde{R}^{-1} \left( \left( \tilde{R}^2 + 1 \right)^{-n} - 1 \right)}
\ee
This is hardly inversible but fortunately the situation improves for $\tilde{R}\gg 1$. 
In this regime we have
\bea
f(R) & \simeq & R - \lambda~R_c\equiv R - 2\Lambda(\infty)\lb{fm}\\
F & \simeq & 1\lb{Fm}\\
r & \simeq & -1-\lambda \tilde{R}^{-1}\lb{rm}\\
m(r)& \simeq & \frac{2 n (2 n+1)}{\lambda^{2n}}(-r-1)^{2n+1}    
                    \simeq 2n(2n+1)\frac{ \lambda}{ \tilde{R}^{2n+1} }\approx 0,~~~~m>0\lb{mSt}\\ 
M^2 & \simeq & \frac{1}{3 F'} = \frac{R}{3m F} \gg R  ~.\lb{M} 
\eea
It is clear that when $f(R)$ satisfies (\ref{fm}), $x_2+x_3$ comes close to zero 
as $FR-f\simeq 2\Lambda(\infty)\ll 6H^2 $.
We see that $M^2$, the scalaron mass introduced in \cite{S80}, becomes very large 
in the past. As noted in \cite{S07}, one should find a mechanism to avoid  $M^2$ 
becoming too large in the early universe. We get also strong oscillations $\delta R$ 
in the past and we illustrate this with the oscillations induced in the behaviour 
of $w_{\rm eff}$ on high redshifts already during the matter phase as displayed on 
Figure 1a. These oscillations could lead to overproduction of scalarons in the very early 
universe \cite{S07}. While the cosmic background today corresponds to $\tilde{R}\sim 1$, 
the regime $\tilde{R}\gg 1$, eqs.(\ref{fm})-(\ref{M}), is quickly reached as we go back 
in time. In this regime $r\approx -1$ and $m\approx 0$ so that a viable matter phase is 
possible followed by a late-time accelerated stage. We have also in this regime $F\approx 1$ 
to very high accuracy. One finds that in this model accelerated expansion starts at 
$z_a\simeq (0.5-1)$ depending on the value of $\Omega_{m,0}$, in agreement with observations 
\cite{MPP07}.

In order to be viable, this theory should not be in conflict with gravity constraints. 
On scales satisfying today $k\ll 2\pi M$, one will not feel the presence of a ``fifth 
force'' caused by the scalar part of the gravitational interaction. 
Hence provided the scalaron mass $M$ corresponding to some experiment is large enough, 
the ``fifth force'' will only be felt on scales that are much smaller than those probed 
by this experiment. 
In particular one will not see it in a Cavendish-type experiment on scales larger 
than $5\times 10^{-2}$cm provided $n\ge 1$, and this range can be increased to even 
lower scales for larger $n$ \cite{S07}. 
We refer the interested reader to the reference \cite{S07} for a thorough discussion of 
the properties of this model and of its potential problems. 

The evolution and growth of matter perturbations can also be seen as an experiment 
probing gravity. In this case the scalaron mass can be much smaller so that deviations 
from GR and the appearance of a ``fifth force''can be felt on larger scales and even 
on cosmic scales. We will consider this in more detail in the next section.  

%%%%%%%%%%%%%%%%%%%%%%%%%%%%%%%%%%%%%%%%%%%%%%%%%%%%%%%%%%%%%%%%%%%%%%%%%
%%%%%%%%%%%%%%%%%%%%%%%%%%%%%%%%%%%%%%%%%%%%%%%%%%%%%%%%%%%%%%%%%%%%%%%%%
\begin{figure}
\begin{center}
\includegraphics[scale=.5]{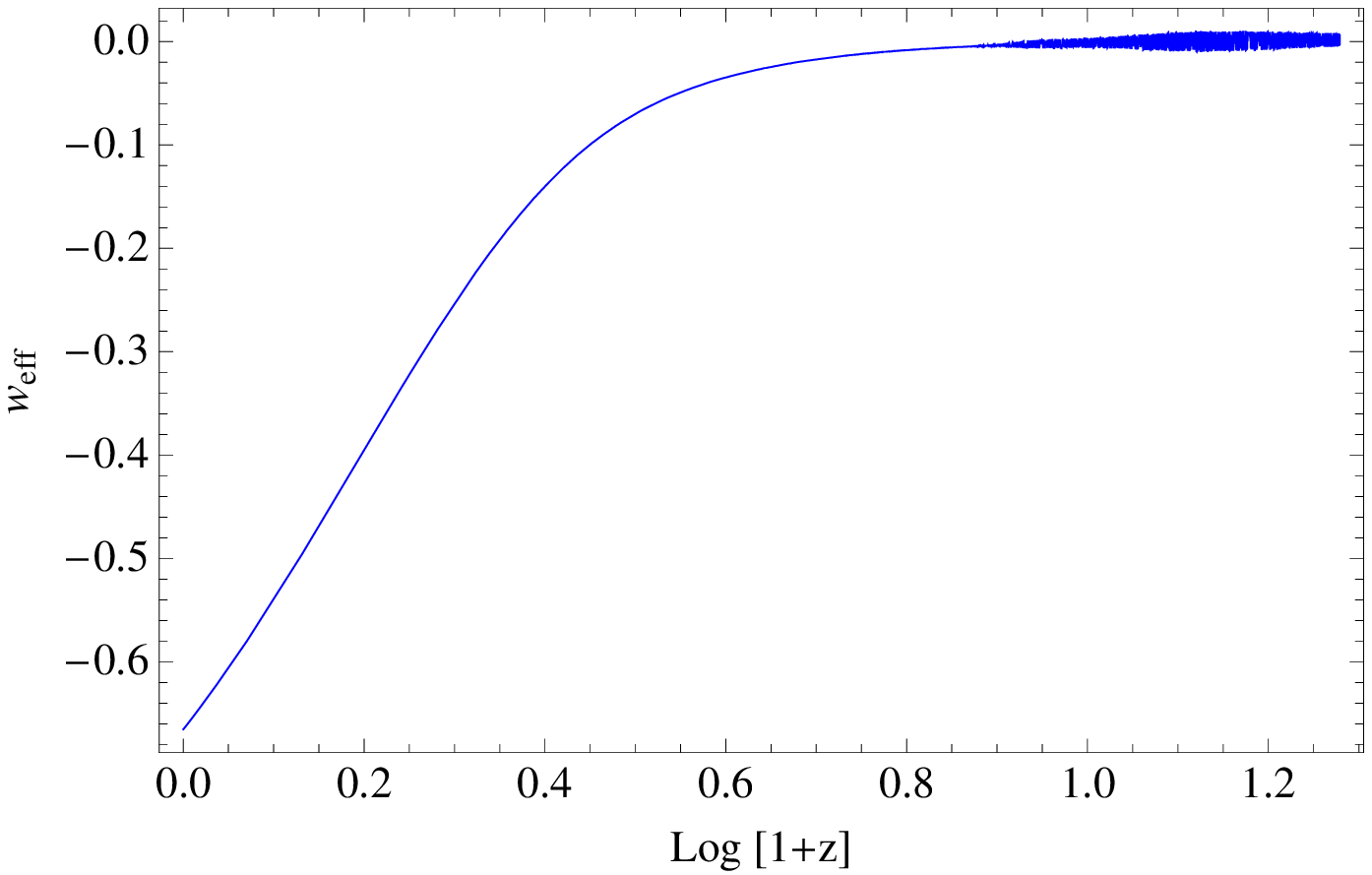} \includegraphics[scale=.53]{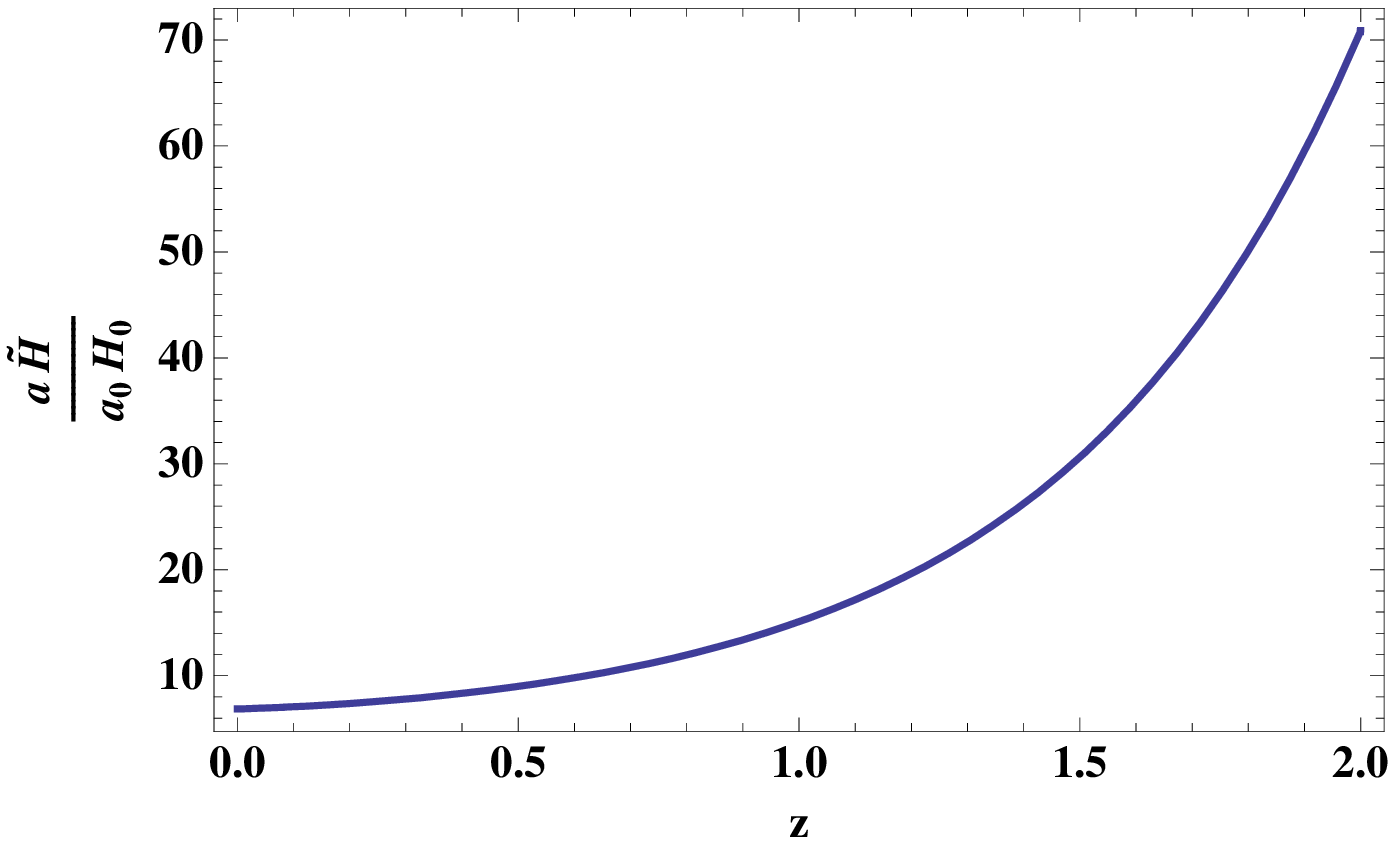}
\caption{ a) On the left panel the effective (total) equation of state parameter 
$w_{eff}$ is shown for $n=2$. We see that large oscillations appear in the past, 
already in the matter-dominated stage at rather low redshifts. ~b) On the right panel, 
the behaviour of the quantity $\frac{a\tilde{H}}{a_0~H_0}$ is displayed. It is seen that 
some subhorizon scales satisfying $k\ll a\tilde{H}$ in the past will gradually shift 
to the regime $k\gg a\tilde{H}$ as the universe expands. In this asymptotic regime, 
the scalaron corresponding to the cosmic backgroung curvature is nearly massless and 
a fifth-force is felt on these cosmic scales. Note that today $\tilde{H}_0\sim 7~H_0$.} 
\end{center}
\end{figure}
%%%%%%%%%%%%%%%%%%%%%%%%%%%%%%%%%%%%%%%%%%%%%%%%%%%%%%%%%%%%%%%%%%%%%%%%%
%%%%%%%%%%%%%%%%%%%%%%%%%%%%%%%%%%%%%%%%%%%%%%%%%%%%%%%%%%%%%%%%%%%%%%%%%
%%%%%%%%%%%%%%%%%%%%%%%%%%%%%%%%%%%%%%%%%%%%%%%%%%%%%%%%%%%%%%%%%%%%%%%%%%%%%%%%%%%%%%%%%%%
%%%%%%%%%%%%%%%%%%%%%%%%%%%%%%%%%%%%%%%%%%%%%%%%%%%%%%%%%%%%%%%%%%%%%%%%%%%%%%%%%%%%%%%%%%%
%%%%%%%%%%%%%%%%%%%%%%%%%%%%%%%%%%%%%%%%%%%%%%%%%%%%%%%%%%%%%%%%%%%%%%%%%%%%%%%%%%%%%%%%%%%
\section{Linear growth of matter perturbations}

In a way analogous to scalar-tensor DE models, the equation governing the growth 
of matter perturbations on subhorizon scales is of the form \cite{BEPS00}
\be
{\ddot \delta_m} + 2H {\dot \delta_m} - 4\pi G_{\rm eff}~\rho_m~\delta_m = 0~.\label{del}
\ee
The expression for $G_{\rm eff}$ in $f(R)$ models is given below, see eq.(42). Many DE models 
outside GR have a modified equation of this type describing the growth of matter perturbations. 
For the models where it is valid, this equation includes all perturbations, the assumption 
is that we are in a regime deep inside the Hubble radius where the leading terms result in 
\eqref{del}. One can also have in principle more elaborate DE models yielding a different 
equation (see e.g. \cite{KS07}). It is convenient to rewrite the corresponding equation 
satisfied by the quantity $f=\frac{d \ln \delta}{d \ln a}$\footnote{The reader should 
not confuse this quantity with the quantity $f(R)$, we believe such confusion will be 
easily avoided from the context where they appear.}\cite{PG08} 
\be
\frac{df}{dN} + f^2 + \frac{1}{2}\left( 1-\frac{d\ln \Omega_m}{dN} \right) f = 
                \frac{3}{2} \frac{G_{\rm eff}}{G_*}~\Omega_m \lb{f}~,
\ee
where $\Omega_m$ is defined from (\ref{OmA}) with $A=1$, and \cite{T07}
\be
G_{\rm eff} = \frac{G_*}{F}~\frac{ 1+ 4\frac{k^2 F'}{a^2 F} }{ 1+ 3\frac{k^2 F'}{a^2 F} }~.\lb{Geff1}
\ee

The r.h.s. of (\ref{f}) is given by the expression $4\pi G_{eff}\frac{\rho_m}{H^2}$ 
which obviously does not depend on (the constant) $A$ so that the evolution of $f$ 
does not depend on $A$ either.   
If we write this r.h.s. using the definition (29) $\Omega_m = \frac{8 \pi G_* \rho_m}{3A H^2}$ 
with $A\ne 1$, we will have $4\pi G_{eff}\frac{\rho_m}{H^2} = \frac{3}{2}~A ~\frac{G_{eff}}{G_*}\Omega_m$ 
and we have in particular $\frac{3}{2}~A~\frac{ G_{\rm eff} }{ G_* } \Omega_m = 
\frac{3}{2} \frac{1+ 4 \frac{k^2 F'}{a^2 F} }{ 1+ 3 \frac{k^2 F'}{a^2 F} }~
\tilde{\Omega}_m$, with $\tilde{\Omega}_m$ defined in eq.(11).
When gravity is described by GR eq.(\ref{f}) reduces to eq.(B7) given in \cite{WS98}.

A crucial point in this model is that $G_{\rm eff}$ is scale dependent 
\be
G_{\rm eff} = G_{\rm eff}(z,k)~.
\ee
In other words, the driving force in eq.(\ref{f}) introduces a scale dependence 
in the growth of perturbations. 
%%%%%%%%%%%%%%%%%%%%%%%%%%%%%%%%%%%%%%%%%%%%%%%%%%%%%%%%%%
%%%%%%%%%%%%%%%%%%%%%%%%%%%%%%%%%%%%%%%%%%%%%%%%%%%%%%%%%%
The spatial variation on cosmic scales of $G_{eff}$ increases the growth of matter 
perturbations as is seen very clearly from Figure 2a. 
As this increase is rather tightly constrained by observations, so will the model 
parameter $n$ on which this increase depends \cite{S07}. We will return in more details 
to this important point below.   

%%%%%%%%%%%%%%%%%%%%%%%%%%%%%%%%%%%%%%%%%%%%%%%%%%%%%%%%%%
%%%%%%%%%%%%%%%%%%%%%%%%%%%%%%%%%%%%%%%%%%%%%%%%%%%%%%%%%%
We can rewrite $G_{\rm eff}$ in the suggestive way 
\bea
G_{\rm eff} &=& \frac{G_*}{F}~ \frac{1+ 4 \frac{k^2}{a^2 \tilde{H}^2}}{1+3 \frac{k^2}{a^2 \tilde{H}^2}}\\
&=& \frac{G_*}{F}~\left( 1 + 
         \frac{\frac{k^2}{a^2 \tilde{H}^2}}{1 + 3 \frac{k^2}{a^2 \tilde{H}^2}}\right)~,\lb{Geff2}
\eea
where 
\be
\tilde{H}^2 = \frac{F}{F'}\equiv 3F M^2 = \frac{R}{m}~.\lb{MHt}
\ee
In equation (\ref{f}), $G_{\rm eff}$ depends on the cosmic curvature $R$ taken from (\ref{R}).  
In the limit ${\tilde R}\gg 1$ satisfied by the cosmic curvature 
$R$ at high redshifts, we have
\be
G_{\rm eff} = G_* ~\left( 1 + \frac{1}{3}~
         \frac{\frac{k^2}{a^2 M^2}}{1 + \frac{k^2}{a^2 M^2}}\right)~~~~~~~~~~~~~~~~~~~~~~~
{\tilde R}\gg 1~.\lb{Geffm}
\ee
%%%%%%%%%%%%%%%%%%%%%%%%%%%%%%%%%%%%%%%%%%%%%
%%%%%%%%%%%%%%%%%%%%%%%%%%%%%%%%%%%%%%%%%%%%%   
Taking the (inverse) Fourier transform of (\ref{Geffm}) it is straightforward to recover 
the corresponding gravitational potential per unit mass $V(r)$. 
Remembering that a potential $\propto \frac{1}{r}$ in real space yields a $k^{-2}$ term in 
Fourier space, we recognize in \eqref{Geffm} the gravitational potential in real space 
(per unit mass)
\be
V(r) = - \frac{G_*}{r}~\left( 1 + \frac{1}{3}~e^{-Mr}\right)~.\lb{Vr}
\ee
%%%%%%%%%%%%%%%%%%%%%%%%%%%%%%%%%%%%%%%%%%%%%
%%%%%%%%%%%%%%%%%%%%%%%%%%%%%%%%%%%%%%%%%%%%%
The quantity $\tilde{H}$ (and $M(R)$) can become small enough with the universe 
expansion, see Figure 1b, so that some cosmic subhorizon scales can feel a significant 
fifth-force. As the universe expands, $\tilde{H}$ is rapidly decreasing so that these 
deviations are felt in Poisson's equation on ever increasing scales. While in the past 
only scales very much smaller than those corresponding to cosmic scales today could feel 
deviations from GR, today $\tilde{H}_0\sim H_0$ and deviations can be felt on essentially 
all subhorizon scales. 

We can distinguish two basic asymptotic regimes 
\bea
G_{\rm eff} &=& \frac{4}{3} \frac{G_*}{F}~~~~~~~~~~~~~~~~~~~~~~~~k\gg a \tilde{H}~,\lb{as1}\\
&=& \frac{G_*}{F}~~~~~~~~~~~~~~~~~~~~~~~~~~k\ll a \tilde{H}~.\lb{as}
\eea
The quantity $a \tilde{H}$ is rapidly increasing in the past. Hence some cosmic subhorizon 
scales ($k\gg a H$) will satisfy $k \ll a \tilde{H}$ in earlier times (large redshifts) and 
switch into the regime $k \gg a \tilde{H}$ on lower redshifts as the universe expands, see 
Figure 1b. For cosmic scales that are in this regime (\ref{as1}), the scalaron mass appears 
negligible and a fifth force does appear which results in the factor $\frac{4}{3}$. 

An accurate evolution of $f$ or $\delta_m$ for arbitrary r.h.s. of eq.(\ref{f}) 
requires numerical calculations. But it is possible to find analytically the 
solutions in the two asymptotic regimes (\ref{as1},\ref{as}) during the matter stage when 
$\Omega_m\approx 1$ and $F\approx 1$. This regime is still valid until 
low redshifts. When $C\equiv \frac{G_{\rm eff}}{G_*} \Omega_m =$ constant we get 
constant growing mode solutions $f=p$ with (see e.g.\cite{GP08}) 
\be
p = \frac{1}{4}\left(-1+\sqrt{1+24C}\right)~.
\ee
A similar result can also be obtained in the framework of chameleon models (see e.g. \cite{BVD04}) 
and inside GR (see e.g. \cite{LP06}). 
We have $C\approx 1$ for $k\ll a \tilde{H}$ and $C\approx \frac{4}{3}$ 
for $k\gg a \tilde{H}$, so that we obtain \cite{S07}
\bea
\delta_m & \propto & a~~~~~~~~~~~~~~~~~~~~~~~~~~~~~~~~~~~~~k\ll a \tilde{H}\lb{delas}\\ 
\delta_m & \propto & a^{\frac{\sqrt{33}-1}{4}}~~~~~~~~~~~~~~~~~~~~~~~~~~~~~~~k\gg a \tilde{H}~.\lb{delas1}
\eea 
For subhorizon scales that go from the first into the second regime, this can yield 
a scale-dependent increase (see Figure 2a) in the growth of matter perturbations. 
Note that this increase takes place on small cosmic scales.
%%%%%%%%%%%%%%%%%%%%%%%%%%%%%%%%%%%%%%%%%%%%%%%%%%%%%%%%%%%%%%%%%%%%%%%%%
%%%%%%%%%%%%%%%%%%%%%%%%%%%%%%%%%%%%%%%%%%%%%%%%%%%%%%%%%%%%%%%%%%%%%%%%%
This increase in the growth of matter perturbations will induce a change of shape 
of the matter power spectrum $P(k)$ inferred from galaxy surveys with a subsequent 
change of its spectral index $n_s^{gal}$, $P(k)\propto k^{n_s^{gal}}$. 
This can then be compared with the spectral index $n_s^{CMB}$ derived from the CMB 
anisotropy data resulting in a possible discrepancy between both spectral indices. 
No significant discrepancy between the two values is allowed by present 
observations and we have the conservative bound (see e.g. \cite{TES06})
\be
n_s^{gal} - n_s^{CMB} < 0.05\lb{deln1}
\ee 
So this difference is rather tightly constrained and so is the model 
parameter $n$ on which it depends.
This results in the constraint $n\ge 2$. 
Because increasing the model parameter $n$ will decrease this discrepancy 
according to the analytical estimate \cite{S07}
\be
n_s^{gal} - n_s^{CMB} = \frac{\sqrt{33}-5}{2(3n+2)}~,\lb{deln2}
\ee 
which is accurate for a wide range of $n$ values (only for large values one 
should resort to a more accurate numerical estimate), (\ref{deln1},\ref{deln2}) 
result in the constraint $n\ge 2$.
%%%%%%%%%%%%%%%%%%%%%%%%%%%%%%%%%%%%%%%%%%%%%%%%%%%%%%%%%%%%%%%%%%%%%%%%%
%%%%%%%%%%%%%%%%%%%%%%%%%%%%%%%%%%%%%%%%%%%%%%%%%%%%%%%%%%%%%%%%%%%%%%%%%

Let us emphasize again at this point the differences with scalar-tensor DE models. 
In scalar-tensor DE models $G_{\rm eff}$ does not depend on ${\bf \vec k}$ (or 
on ${\bf \vec r}$ in real space), at a given time it is the same $G_{\rm eff}$ that 
enters the equation for the growth of perturbations and the gravitational constant 
measured in a Cavendish-type experiment. These DE models have further a negligible 
dilaton mass but they can comply with solar-system constraints if they satisfy 
$\omega_{BD,0}>4\times 10^4$, where $\omega_{BD,0}$ is the value today of the 
Brans-Dicke parameter, which yields $\gamma_{PN}\approx 1$ \cite{BEPS00},\cite{EP01}.
Indeed for scalar-tensor models one has for the Post-Newtonian parameter today 
$\gamma_{PN}=\frac{\omega_{BD,0}+1}{\omega_{BD,0}+2}$ and solar-system constraints 
point at $\gamma_{PN}\approx 1$ close to its GR value $\gamma_{PN}=1$. As $f(R)$ models 
correspond to scalar-tensor models with $\omega_{BD}=0$ they have 
$\gamma_{PN}=\frac{1}{2}$ which requires a high scalaron mass to comply with 
solar-system constraints. 

In asymptotically stable scalar-tensor DE models the matter perturbations grow slowlier 
than $a$ in the matter stage and this growth is scale-independent.  
%%%%%%%%%%%%%%%%%%%%%%%%%%%%%%%%%%%%%%%%%%%%%%%%%%%%%%%%%%%%%%%%%%%%%%%%%
%%%%%%%%%%%%%%%%%%%%%%%%%%%%%%%%%%%%%%%%%%%%%%%%%%%%%%%%%%%%%%%%%%%%%%%%%
\begin{figure}
\begin{center}
\includegraphics[scale=.7]{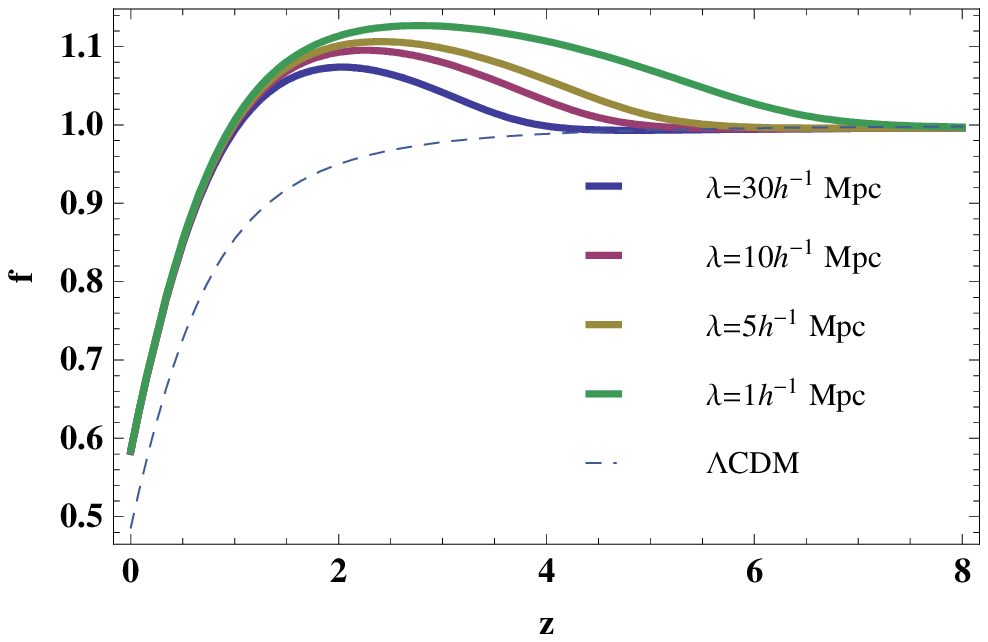} \includegraphics[scale=.7]{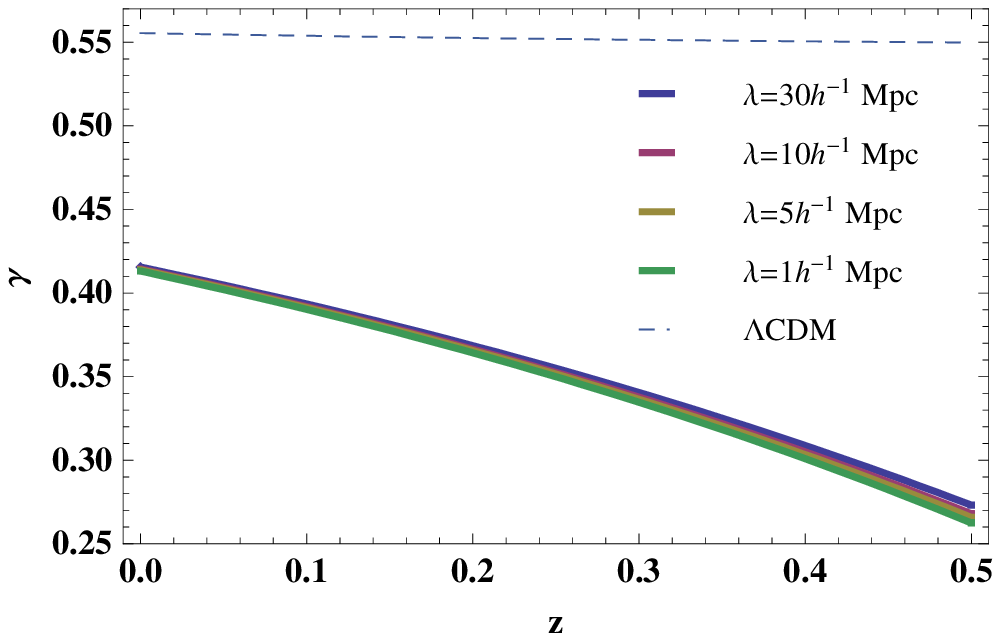} \includegraphics[scale=.7]{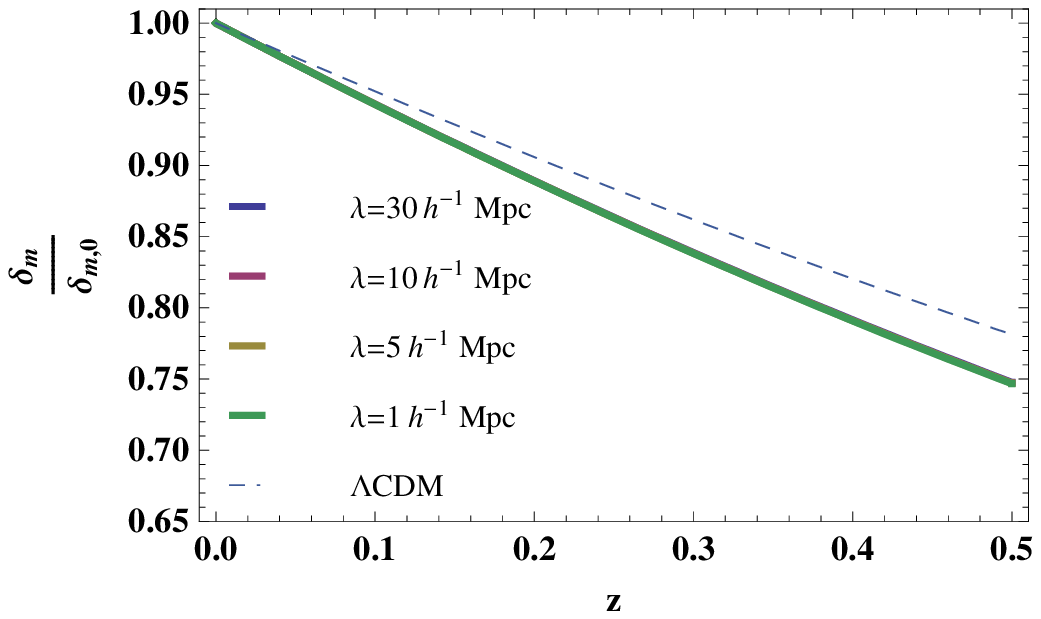}
\caption{a) On the left, the growth factor $f$ is shown for the model with $n=2$ 
and the dimensionless model parameter $\lambda$ has the value $\lambda=0.94$. 
We see the scale dependent behaviour of $f$. On low redshifts all the cosmic 
scales displayed here are in the regime \eqref{as1},and matter perturbations obey 
\eqref{delas1} in the matter-dominated stage. A significant fifth-force is felt 
in the deviation of Poisson's equation for matter perturbations on these cosmic 
scales. Note that on higher 
redshifts these scales will eventually be in the opposite regime \eqref{as}, so 
that perturbations follow the evolution \eqref{delas}. This is the origin of the bumps 
seen on the figure. b) On the right, the quantity $\gamma(z)$ is 
displayed for the same model parameters as on the left panel. We find a low value 
for the growth parameter $\gamma_0$, substantially lower than in $\Lambda$CDM. 
It is seen that the dispersion of $\gamma(z)$ is very small, while that of $\gamma_0$ 
is negigible. We find also a value for $\gamma'_0\equiv \frac{d\gamma}{dz}(z=0)$, 
$\gamma'_0\sim 0.2$ which is much larger than the value found for DE models inside GR.
c) On the bottom figure, the quantity $\frac{\delta_m(z)}{\delta_{m,0}}$ is 
displayed up to $z=0.5$. This quantity does not depend explicitly on the background 
quantity $\Omega_m$. At $z=0.5$, a difference of about $6\%$ is found with $\Lambda$CDM.}

\end{center}
\end{figure}
%%%%%%%%%%%%%%%%%%%%%%%%%%%%%%%%%%%%%%%%%%%%%%%%%%%%%%%%%%%%%%%%%%%%%%%%%
%%%%%%%%%%%%%%%%%%%%%%%%%%%%%%%%%%%%%%%%%%%%%%%%%%%%%%%%%%%%%%%%%%%%%%%%%

The growth of matter perturbations could provide an efficient way to discriminate 
between modified gravity DE models and DE models inside GR. 
One can characterize the growth of matter perturbations on small redshifts using 
the quantity $\gamma(z)$
\be
f=\Omega_m(z)^{\gamma(z)}~.  
\ee
In the pioneering papers on this approach $\gamma$ was taken constant \cite{P84}. But 
it is important to realize that $\gamma(z)$ is a function of $z$ which is generically 
not constant, and that its variation can contain crucial information about the 
underlying model. As was shown earlier, for a wide class of models inside 
GR one has $|\gamma'_0\equiv \frac{d\gamma}{dz}(z=0)|\lesssim 0.02$ so that $\gamma(z)$ 
is approximately constant \cite{PG08}. 
In $\Lambda$CDM we have $\gamma_0\approx 0.55$ (with a slight dependence on 
$\Omega_{m,0}$) and $\gamma'_0\simeq -0.015$.
However, $\gamma'_0$ can be significantly larger 
in models outside GR \cite{GP08}. From the growth of matter perturbations we can calculate 
$\gamma(z)$ and $\frac{d\gamma}{dz}$  and in particular their value at $z=0$. 
As we can see from Figure 2a, we find that $f$ is scale-independent on low 
redshifts $z\lesssim 0.5$. This means that the growth of matter perturbations is the same 
on all relevant scales in this redshift interval. This is easily understood as 
${\tilde H}_0\sim H_0$ so that all subhorizon scales, and certainly the relevant scales 
that are deep inside the Hubble radius today, satisfy $k\gg {\tilde H}_0$ (we take $a_0=1$).
Some scale dependence can appear in the growth of matter perturbations \emph{today} for 
scales that are so large that the original equation (\ref{f}) (or (\ref{del})) no longer 
provides an accurate approximation. 
Therefore the quantity $\gamma(z)$ is essentially scale independent too on low redshifts 
$z\le 0.3$ in our model. On higher redshifts some restricted dispersion appears 
which could be another signature of this model, for example at $z=0.5$ we can have 
a difference $\Delta\gamma\sim 0.04$ between various scales, see Figure 2b.
On even higher redshifts the growth is of course scale-dependent and we can see from Figures 2a 
that the relative increase is damped with increasing $n$ in accordance with earlier analytical 
estimates. As mentioned earlier, this increase is constrained by the observations and cannot be 
large. 
%%%%%%%%%%%%%%%%%%%%%%%%%%%%%%%%%%%%%%%%%%%%%%%%%%%%%%%%%%%%%%%%%%%%%%%%%%%%%%%%%%%%%
%%%%%%%%%%%%%%%%%%%%%%%%%%%%%%%%%%%%%%%%%%%%%%%%%%%%%%%%%%%%%%%%%%%%%%%%%%%%%%%%%%%%%
\begin{table}
\begin{center}
\begin{tabular}{| c || c | c | c | c | c | c | c | c | c | c |}
\hline
$\Omega_{m,0}$ & 0.322 & 0.315 &  0.302 & 0.289 & 0.273 & 0.263 & 0.254 & 0.245 & 0.238 & 0.227\\
\hline
$\gamma_0$ & 0.396 & 0.399 & 0.404 & 0.409 & 0.415 & 0.419 & 0.422 & 0.425 & 0.428 & 0.432\\
$\gamma'_0$ & -0.253 & -0.246 & -0.234 & -0.224 & -0.210 & -0.202 & -0.195 & -0.189 & -0.183 & -0.175\\
$z_a$ & 0.654 & 0.673 & 0.711 & 0.747 & 0.798 & 0.830 & 0.862 & 0.893 & 0.922 & 0.965\\
$\frac{\delta_m(z=0.5)}{\delta_{m,0}}$ & 0.731 & 0.733 & 0.737 & 0.742 & 0.747 & 0.751 & 0.754 & 0.758 & 0.761 & 0.765\\
\hline
\hline
\end{tabular}
\end{center}
\caption{This table summarizes the value of the growth parameters $\gamma_0$, $\gamma'_0$ and 
of $z_a$, the redshift when accelerated expansion starts. All these values correspond to $n=2$ and 
$\lambda=0.94$.}
\end{table}
%%%%%%%%%%%%%%%%%%%%%%%%%%%%%%%%%%%%%%%%%%%%%%%%%%%%%%%%%%%%%%%%%%%%%%%%%%%%%%%%%%%%%
%%%%%%%%%%%%%%%%%%%%%%%%%%%%%%%%%%%%%%%%%%%%%%%%%%%%%%%%%%%%%%%%%%%%%%%%%%%%%%%%%%%%%
We find further that in this $f(R)$ model $\gamma_0\equiv \gamma(z=0)\approx 0.41$ which 
is much lower than in $\Lambda$CDM where $\gamma_0\approx 0.55$ (and $\gamma'_0\approx -0.015$). 
This is an interesting property which clearly allows to discriminate this model from $\Lambda$CDM. 
It is also significantly lower than the value found in some scalar-tensor DE models. This value 
seems essentially independent of the model parameter $n$. 
So a measurement of $\gamma_0$ could allow to discriminate this model from $\Lambda$CDM, but 
also possibly from other modified gravity DE models. As $\gamma_0$ is scale independent this is 
a clear signature of this $f(R)$ model. 

These conclusions remain the same if we let $\Omega_{m,0}$ vary in the viable range of 
$\Omega_{m,0}$ values. As we can see on Figures 3, there is some variation of  $\gamma_0$ 
and $\gamma'_0$ in function of $\Omega_{m,0}$. We get $\gamma_0\simeq 0.4$ for $\Omega_{m,0}=0.32$ 
and $\gamma_0\simeq 0.43$ for $\Omega_{m,0}=0.23$. 
As for $\gamma'_0$ we have $\gamma'_0\simeq -0.18$ for $\Omega_{m,0}=0.23$ and $\gamma'_0\simeq -0.25$ 
for $\Omega_{m,0}=0.32$. So a low matter density universe brings the quantities $\gamma_0$ and 
$\gamma'_0$ closer to their values in $\Lambda$CDM but they still remain far from them. 
Such low values for $\gamma_0$ seem in trouble with present observations (see e.g. \cite{YSH08})
but an appropriate analysis of the data should relax the assumption $\gamma$= constant. 
It is likely that such a low value will remain in tension with observations. 

%%%%%%%%%%%%%%%%%%%%%%%%%%%%%%%%%%%%%%%%%%%%%%%%%%%%%%%%%%%%%%%%%%%%%%%%%%
%%%%%%%%%%%%%%%%%%%%%%%%%%%%%%%%%%%%%%%%%%%%%%%%%%%%%%%%%%%%%%%%%%%%%%%%%%
\begin{figure}
\begin{center}
\includegraphics[scale=.7]{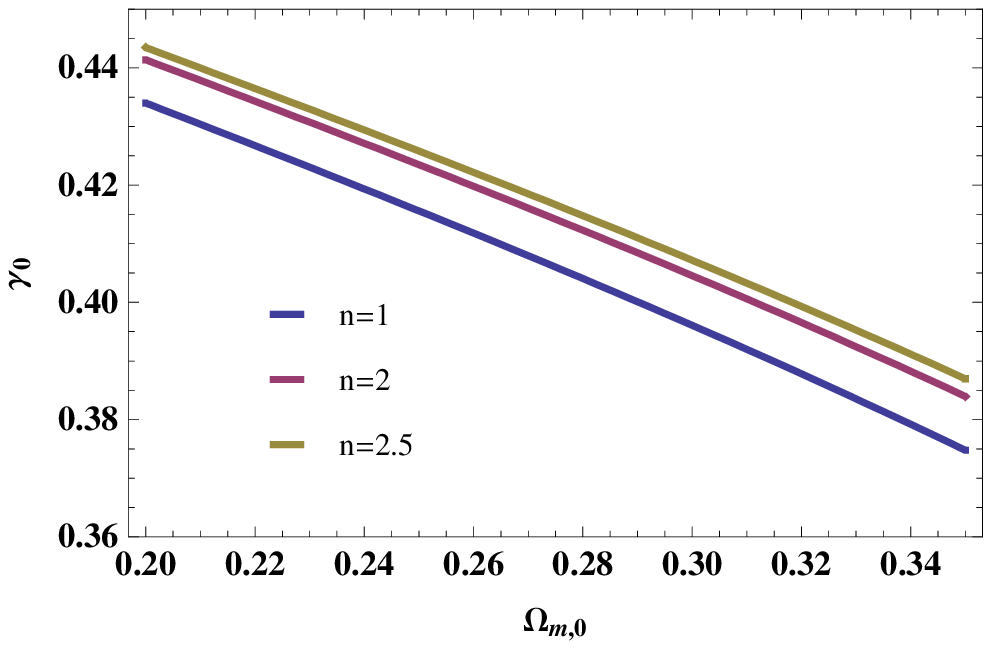} \includegraphics[scale=.7]{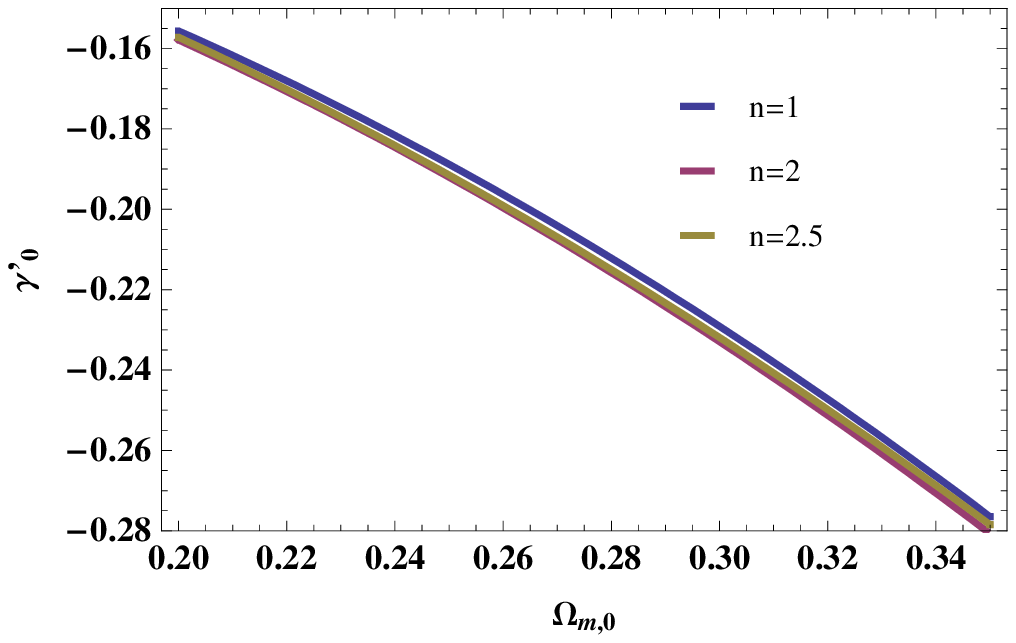}
\caption{ a) On the left, the parameter $\gamma_0$ is displayed in function of 
the cosmological parameter $\Omega_{m,0}$ for different values of the model 
parameter $n$. For given $n$, we see some variation with the highest value 
obtained for a low matter density universe. Still all these values remain 
clearly distinguishable from their values in $\Lambda$CDM. b) On the right, 
the parameter $\gamma'_0$. We note that very large values are obtained for 
all models.}
\end{center}
\end{figure}
%%%%%%%%%%%%%%%%%%%%%%%%%%%%%%%%%%%%%%%%%%%%%%%%%%%%%%%%%%%%%%%%%%%%%%%%%%
%%%%%%%%%%%%%%%%%%%%%%%%%%%%%%%%%%%%%%%%%%%%%%%%%%%%%%%%%%%%%%%%%%%%%%%%%%

Finally we have here again an example of a modified gravity model where 
$\frac{d\gamma}{dz}(z=0)\equiv \gamma'_0\ne 0$ and actually large.
We obtain the high value $\gamma'_0\approx -0.14$ which is largely outside the range 
$|\gamma'_0|\le 0.02$ found for DE models inside GR. It is also much higher than the 
value found for some scalar-tensor DE models. So we have here again a characteristic 
signature of our model which clearly differentiates it both from $\Lambda$CDM and DE 
models inside GR, but possibly also from other DE models outside GR. 
Like for $\gamma_0$, the scale independence of $\gamma'_0$ makes it a clear signature 
of the model. We have checked that $\gamma'_0$ obeys the constraint
\be
\gamma'_0~=~\left[\ln \Omega_{m,0}^{-1}\right]^{-1}~\left[-\Omega_{m,0}^{\gamma_0} - 
                             3(\gamma_0-\frac{1}{2})w_{\rm eff,0} +
       \frac{3}{2} \frac{G_{\rm eff}(R_0)}{G_*}\Omega_{m,0}^{1-\gamma_0}-\frac{1}{2}\right]~,
\ee
where we have to include the factor $\frac{G_{\rm eff}(R_0)}{G_*}\simeq \frac{4}{3 F_0}\ne 1$.
As for the scalar-tensor models considered earlier, we obtain a nearly linear behaviour on 
low redshifts $z\le 0.3$.

In conclusion we find that the growth of matter perturbations on small redshifts provides 
a powerful constraint on our $f(R)$ model. We find low values for the parameter 
$\gamma_0$ with $\gamma_0\approx 0.4$, and high values for $\gamma'_0$ with 
$\gamma'_0\simeq -0.2$. 
These values are not scale dependent and provide therefore a clear signature. 
The growth  parameters $\gamma_0$ and $\gamma'_0$ are mainly affected by the cosmological 
parameter $\Omega_{m,0}$, in particular $\gamma'_0$ is less negative for low $\Omega_{m,0}$ 
values. But in all cases, characteristic values very far from $\Lambda$CDM and from all DE 
models inside GR are found.

There are still very large uncertainties on the quantity $f(z)$ or on $\beta\equiv \frac{f}{b}$ 
where $b$ is the bias factor (see e.g. \cite{YSH08},\cite{NP08}). 
One should further keep in mind that a precise observational determination of both $f(z)$ \emph{and} 
$\Omega_m$ is needed in order to measure $\gamma(z)$ accurately. Hence to get precise values for 
the couple $\gamma_0,\gamma'_0$ we need to determine accurately both $f(z)$ and $\Omega_m(z)$ 
around $z=0$. If future surveys will constrain $\gamma_0,\gamma'_0$ to be close to their 
values for $\Lambda$CDM then the models we have investigated here will be ruled out. 
Though a systematic numerical exploration in the model parameter space is very hard to 
achieve, the model for $n=2$ shows already a very large deviation from $\Lambda$CDM if 
one considers the growth of matter perturbations while this model meets all other 
constraints. So either the model will be ruled out for any value of $n$ or the growth of 
matter perturbations will at least significantly restrict the viable interval in the 
model parameter $n$. 

We conjecture that many if not all viable $f(R)$ models will have similar observational 
signatures. A precise determination of the parameters $\gamma_0$ and $\gamma'_0$ could be 
decisive in the quest for the true DE model especially if it is an $f(R)$ modified gravity 
DE model.  

%%%%%%%%%%%%%%%%%%%%%%%%%%%%%%%%%%%%%%%%%%%%%%%%%%%%%%%%%%%%%%%%%%%%%%%%%%%%%%%%%%%%%%%%%%%
%%%%%%%%%%%%%%%%%%%%%%%%%%%%%%%%%%%%%%%%%%%%%%%%%%%%%%%%%%%%%%%%%%%%%%%%%%%%%%%%%%%%%%%%%%%
%%%%%%%%%%%%%%%%%%%%%%%%%%%%%%%%%%%%%%%%%%%%%%%%%%%%%%%%%%%%%%%%%%%%%%%%%%%%%%%%%%%%%%%%%%%

%\section*{Acknowledgments}

\end{document}